\documentclass[twocolumn]{aastex631}

\begin{document}

\title{Isolated Black Holes as Potential PeVatrons and Ultrahigh-energy Gamma-ray Sources}

\correspondingauthor{Shigeo S. Kimura}
\email{shigeo@astr.tohoku.ac.jp}

\author[0000-0003-2579-7266]{Shigeo S. Kimura}
\affiliation{Frontier Research Institute for Interdisciplinary Sciences, Tohoku University, Sendai 980-8578, Japan}
\affiliation{Astronomical Institute, Graduate School of Science, Tohoku University, Sendai 980-8578, Japan}

\author[0000-0001-8105-8113]{Kengo Tomida}
\affiliation{Astronomical Institute, Graduate School of Science, Tohoku University, Sendai 980-8578, Japan}

\author[0000-0003-3990-1204]{Masato I.N. Kobayashi}
\affiliation{I. Physikalisches Institut, Universität zu Köln, Zülpicher Str. 77, D-50937 Köln, Germany}

\author[0000-0002-9712-3589]{Koki Kin}
\affiliation{Astronomical Institute, Graduate School of Science, Tohoku University, Sendai 980-8578, Japan}

\author[0000-0002-9725-2524]{Bing Zhang}
\affiliation{Nevada Center for Astrophysics, University of Nevada, Las Vegas, NV 89154, USA}
\affiliation{Department of Physics and Astronomy, University of Nevada, Las Vegas, NV 89154, USA}



\begin{abstract}
 The origin of PeV cosmic rays is a long-standing mystery, and ultrahigh-energy gamma-ray observations would play a crucial role in identifying it. Recently, LHAASO reported the discovery of ``dark'' gamma-ray sources that were detected above 100 TeV without any GeV--TeV gamma-ray counterparts. The origins of these dark gamma-ray sources are unknown. We propose isolated black holes (IBHs) wandering in molecular clouds as the origins of PeV cosmic rays and LHAASO dark sources. An IBH accretes surrounding dense gas, which forms a magnetically arrested disk (MAD) around the IBH. Magnetic reconnection in the MAD can accelerate cosmic-ray protons up to PeV energies. Cosmic-ray protons of GeV-TeV energies fall to the IBH, whereas cosmic-ray protons at sub-PeV energies can escape from the MAD, providing PeV CRs into the interstellar medium. The sub-PeV cosmic-ray protons interact with the surrounding molecular clouds, producing TeV-PeV gamma rays without emitting GeV-TeV gamma rays. This scenario can explain the dark sources detected by LHAASO.  Taking into account the IBH and molecular cloud distributions in our Galaxy, we demonstrate that IBHs can provide a significant contribution to the PeV cosmic rays observed on Earth. Future gamma-ray detectors in the southern sky and neutrino detectors would provide a concrete test to our scenario.
\end{abstract}



\section{Introduction} \label{sec:intro}

The origin of high-energy cosmic rays, especially above PeV energies, have been a long-lasting mystery in astrophysics. Recent observations of gamma-ray (Tibet AS$\gamma$ and LHAASO) and neutrinos (IceCube) at TeV-PeV energies provide strong evidence that PeV CR accelerators reside in our Galaxy \citep{2021PhRvL.126n1101A,2023PhRvL.131o1001C,2023Sci...380.1338I}.  Cosmic rays of GeV-TeV energies are believed to originate from supernova remnants (SNRs). This is strongly supported by GeV-TeV gamma-ray observations \citep{Ackermann:2013wqa,2021Ap&SS.366...58S}. However, historical SNRs show a break or cutoff at $E\sim0.1-10$ TeV in their gamma-ray spectra \citep{MAGIC17a,Giuliani:2024fpq}, raising a question whether SNRs can accelerate CRs up to PeV energies.

Black holes (BHs) in our Galaxy are suggested as alternative PeVatrons. \cite{FMK17a} and \cite{2022ApJ...935..159K} suggested Sgr A* as PeV - EeV CR sources, considering that its accretion flows accelerate CRs by stochastic acceleration and magnetic reconnection, respectively. Micro-quasars are also discussed as PeV CR sources. \cite{2020MNRAS.493.3212C} considered PeV CR production in jets of luminous X-ray binaries. \cite{2021ApJ...915...31K} considered PeV CR production in accretion flows in quiescent BH binaries. \cite{IMT17a} suggested stellar-mass isolated BHs (IBHs) formed by binary BH mergers as PeVatrons. \cite{2012MNRAS.427..589B} proposed IBHs with magnetized low angular momentum accretion flows as TeV--PeV leptonic CR sources. These papers considered CR production in jets.

  \begin{figure*}
   \begin{center}
    \includegraphics[width=\linewidth]{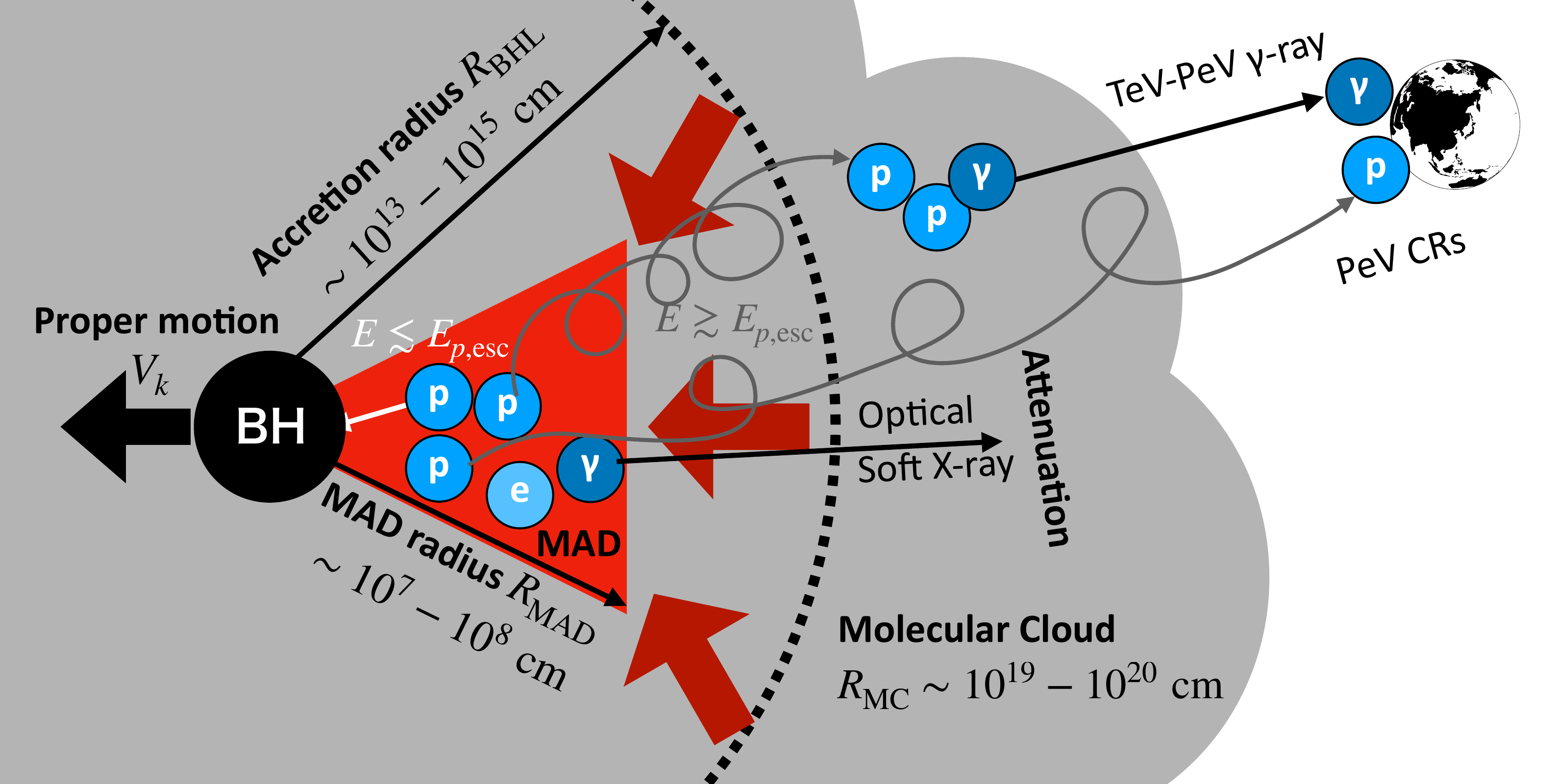}
    \caption{A schematic picture of our scenario. An IBH in a molecular cloud accretes the surrounding gas, forming a MAD. Protons are accelerated in the MAD and high-energy protons can escape from the MAD. Some of these protons interact with the ambient gas, emitting TeV-PeV gamma rays that can explain LHAASO dark sources. The majority of the protons escape from the molecular cloud, which contribute to the PeV CRs observed on Earth. }
    \label{fig:schematic}
   \end{center}
  \end{figure*}

In order to confirm these scenarios, ultrahigh-energy (UHE) gamma-ray observations are crucial. Recently, LHAASO and HAWC reported sub-PeV gamma rays around micro-quasars, which strongly support that stellar-mass BHs in our Galaxy accelerate hadronic cosmic-rays up to multi-PeV energies \citep{HAWC24Natur.634..557A,2024arXiv241008988L}.
In addition, LHAASO identified 43 UHE gamma-ray sources \citep{2024ApJS..271...25C}. This list includes 7 newly discovered ``dark'' gamma-ray sources, from which LHAASO detected gamma rays with soft spectra above 30 TeV without showing signatures of lower-energy gamma-rays of 0.1 -- 2000 GeV. Their gamma-ray spectra likely have peaks around 30 TeV. Such objects had not been reported before, and the origins of these dark sources became a new mystery. 

In this Letter, we propose isolated black holes (IBHs) wandering in molecular clouds as PeVatrons and LHAASO dark sources. Based on stellar evolution theories, $10^8-10^9$ IBHs are expected to be wandering in the interstellar medium (ISM) in our Galaxy \citep[e.g.,][]{2020ApJ...905..121A}, which accrete gas from the ISM. Figure \ref{fig:schematic} indicates a schematic picture of our scenario.
The accretion flows onto these IBHs are considered to be in a magnetically arrested disk (MAD) state \citep[][: see Section \ref{sec:accretion}]{IMT17a,2021ApJ...922L..15K}, where magnetic reconnection can efficiently accelerate non-thermal particles. We show that MADs around IBHs embedded in molecular clouds can accelerate CRs up to PeV energies. The high-energy CRs can escape from MADs, and a fraction of them would interact with ambient molecular clouds. This interaction produces UHE gamma-rays that can explain LHAASO dark sources.  The vast majority of the PeV CRs escape from the molecular clouds, and these CRs can provide a significant contribution to the CRs observed on Earth. Throughout the Letter, we use notation of  $Q_X=Q/10^X$ in cgs units unless otherwise noted.

\section{Accretion flows onto IBHs in Molecular Clouds} \label{sec:accretion}

We consider a stellar-mass IBH wandering in a molecular cloud. The IBH captures the ambient gas with the Bondi-Hoyle-Lyttleton rate, but a fraction of the accreting gas would not reach the vicinity of the IBHs because of mass loss or convective motion \citep{BB99a,2000ApJ...539..809Q}. We introduce a parameter, $\lambda_w$, to take into account the reduction of mass accretion rate. The value of $\lambda_w$ is under debate; It would also depend on efficiencies of kinetic/radiation feedback \citep[e.g.,][]{2017MNRAS.469...62S,2024MNRAS.528.2588O}. We here use $\lambda_w=0.1$ as a reference value, which is consistent with recent general relativistic magnetohydrodynamic (GRMHD) simulations \citep{2024arXiv240911486G,2024arXiv240912359K}. Then, we estimate the mass accretion rate onto IBH as
\begin{eqnarray} \dot{M}_{\bullet}&\approx&\lambda_w\frac{4\pi G^2M^2\mu_{\rm MC}m_pn_{\rm MC}}{(C_{s,\rm eff}^2+V_k^2)^{3/2}}\nonumber\\
&\simeq& 1.1\times10^{14}\lambda_{w,-1}M_1^2n_{\rm MC,2}V_{\rm eff,6.3}^{-3}\rm~g~s^{-1},
\end{eqnarray}
where $G$ is the gravitational constant, $M$ is the IBH mass, $m_p$ is the proton mass, $V_k$ is the relative velocity between the IBH and the molecular gas, $\mu_{\rm MC}=2.3$, $n_{\rm MC}$, and $C_{s,\rm eff}\sim10^6\rm~cm~s^{-1}$ are the mean molecular weight, number density, and the effective sound speed including turbulence velocity dispersion in the molecular gas, respectively, with $V_{\rm eff}=\sqrt{C_{s,\rm eff}^2+V_k^2}$, and $M_1=M_\bullet/10~M_\odot$.
This value is much lower than the Eddington accretion rate; The Eddington ratio is estimated to be 
\begin{eqnarray}
 \dot{m}=\frac{\dot{M}_\bullet c^2}{L_{\rm Edd}} \simeq 7.6\times10^{-5} M_1n_{\rm MC,2}\lambda_{w,-1}V_{\rm eff,6.3}^{-3}.
\end{eqnarray}
With such a low Eddington ratio, we expect formation of hot accretion flows \citep{YN14a}, which carries a magnetic flux in the ISM efficiently owing to the rapid advection.
This causes accumulation of magnetic flux onto the IBH  \citep{2011ApJ...737...94C,2021ApJ...915...31K,2023ApJ...944..182D}. Based on GRMHD simulations, the magnetic flux threading a BH has a saturation value, and a MAD is formed if the magnetic flux threading a BH reaches this value \citep{TNM11a,MTB12a}.  Since the magnetic flux within the Bondi radius is much higher than the saturation flux in a typical ISM environment \citep{IMT17a}, we expect the formation of a MAD around the IBH.
Also, recent GRMHD simulations revealed that a hot accretion flow can reach the MAD state even without the initial net poloidal magnetic field \citep{2020MNRAS.494.3656L}, which also supports the formation of a MAD.

The Eddington ratio of MADs around IBHs in molecular clouds is comparable to those for quiescent X-ray binaries.
\cite{2021ApJ...915...31K} constructed a multi-wavelength emission model considering MADs in quiescent X-ray binaries. This model successfully explains the optical and X-ray data. Assuming that the plasma state of MADs around IBHs are similar to that in quiescent X-ray binaries, we use the same plasma parameters as those in \cite{2021ApJ...915...31K}. Based on the parameterization,
 15\% of the released energy is dissipated, $L_{\rm dis}=\epsilon_{\rm diss}\dot{M}c^2$ with $\epsilon_{\rm diss}=0.15$. Protons and electrons would obtain 70\% and 30\% of the dissipation energy, so that $L_p=(1-f_e)L_{\rm diss}$ and $L_e=f_eL_{\rm diss}$ with $f_e=0.3$. Non-thermal particles would obtain 1/3 of the dissipation energy, $L_{\rm CR}=\epsilon_{\rm NT}L_p$ with $\epsilon_{\rm NT}=0.33$, which leads to a cosmic-ray proton luminosity of
\begin{equation}
 L_{\rm CR}\simeq 4.1\times10^{33}M_1^2n_{\rm MC,2}\lambda_{w,-1}V_{\rm eff,6.3}^{-3}f_{\rm CR,-1.5}\rm~erg~s^{-1},
\end{equation}
where $f_{\rm CR}=\epsilon_{\rm dis}\epsilon_{\rm NT}(1-f_e)$ is the fraction of accretion energy that goes to the CR proton energy. 
The magnetic field strength in the MAD is estimated by assuming a value of plasma beta ($\beta=P_g/P_B$, where $P_g$ and $P_B$ are the gas and magnetic pressure, respectively) in the emission region, which leads to 
\begin{equation}
B\simeq 6.3\times10^{5}n_{\rm MC,2}^{1/2}\lambda_{w,-1}^{1/2}V_{\rm eff,6.3}^{-3/2}\mathcal{R}_1^{-5/4}\alpha_{-0.5}^{-1/2}\beta_{-1}^{-1/2} \rm~G,
\end{equation}
where $\mathcal{R}=R_{\rm MAD}/R_G$, $R_{\rm MAD}$ is the emission radius, $R_G=GM/c^2$ is the gravitational radius, and $\alpha$ is the viscous parameter. We use $\alpha=0.3$ and $\beta=0.1$ as in the case of quiescent X-ray binaries.
Since the MADs have strong and ordered magnetic fields \citep[e.g.,][]{2019ApJ...874..168W}, the plasma beta at the emission region can be lower than unity. This is a distinct feature of MADs from the classical hot accretion flows in which magnetic fields are generated by magnetorotational instability. 

Molecular clouds have a wide range of density structure within them. The majority of their volume has $n_{\rm MC} = 10^2 - 10^3 \rm~cm^{-3}$, while some volumes form filaments and cores with $n_{\rm MC} = 10^4 -10^5 \rm~cm^{-3}$ \citep{2010A&A...518L.102A}.
If relatively massive IBHs ($M\gtrsim30 M_\odot$) are located in such dense environments, the Eddington ratio of the accretion flows are too high to maintain the structure of hot accretion flows, which would likely lead to the formation of geometrically thin accretion disks. In this situation, the transport of the magnetic flux in the ISM is inefficient due to its slow radial velocity  \citep{1994MNRAS.267..235L}, which prevents MADs from forming around IBHs. We focus on the parameter space where we would expect the formation of hot accretion flows around the IBHs, i.e., $\dot{m}<\alpha^2\sim0.1$ \citep[e.g.,][]{xy12,YN14a}\footnote{ We should note that our definition of $\dot{m}$ does not include the radiation efficiency parameter, $\epsilon_{\rm rad}\sim0.1$. This causes that the value of $\dot{m}$ in our paper is 10 times higher than that in other papers that defines the Eddington ratio with $\epsilon_{\rm rad}$.}.

\section{IBHs in Molecular Clouds as PeVatrons}

  \begin{figure}
   \begin{center}
    \includegraphics[width=\linewidth]{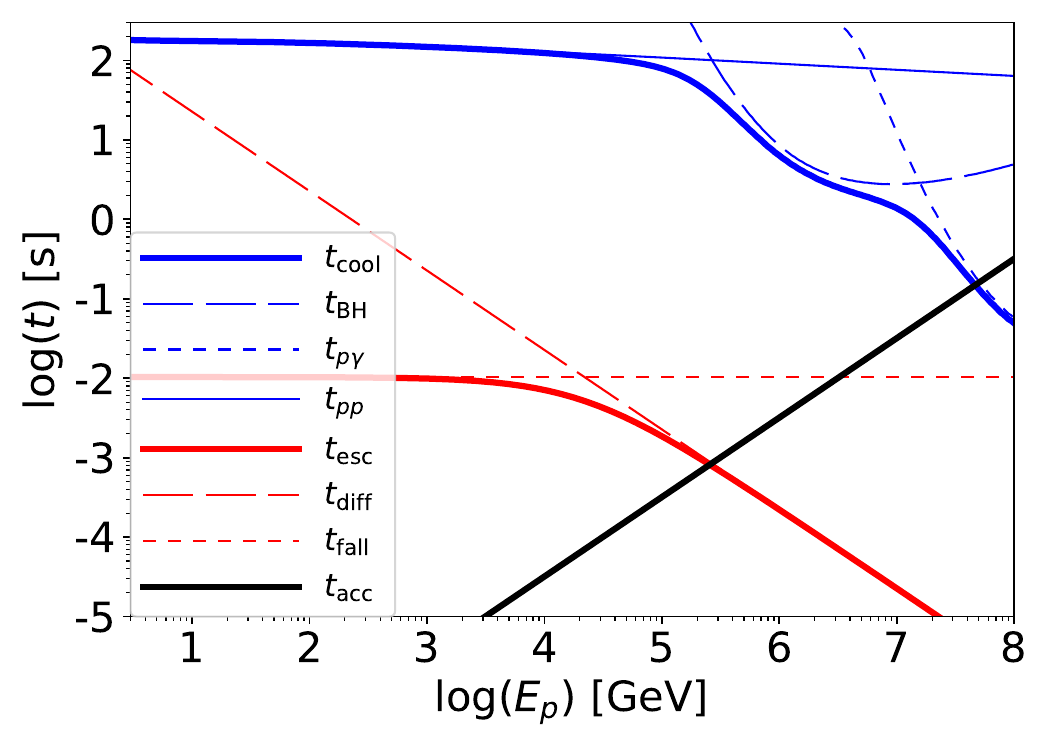}
    \includegraphics[width=\linewidth]{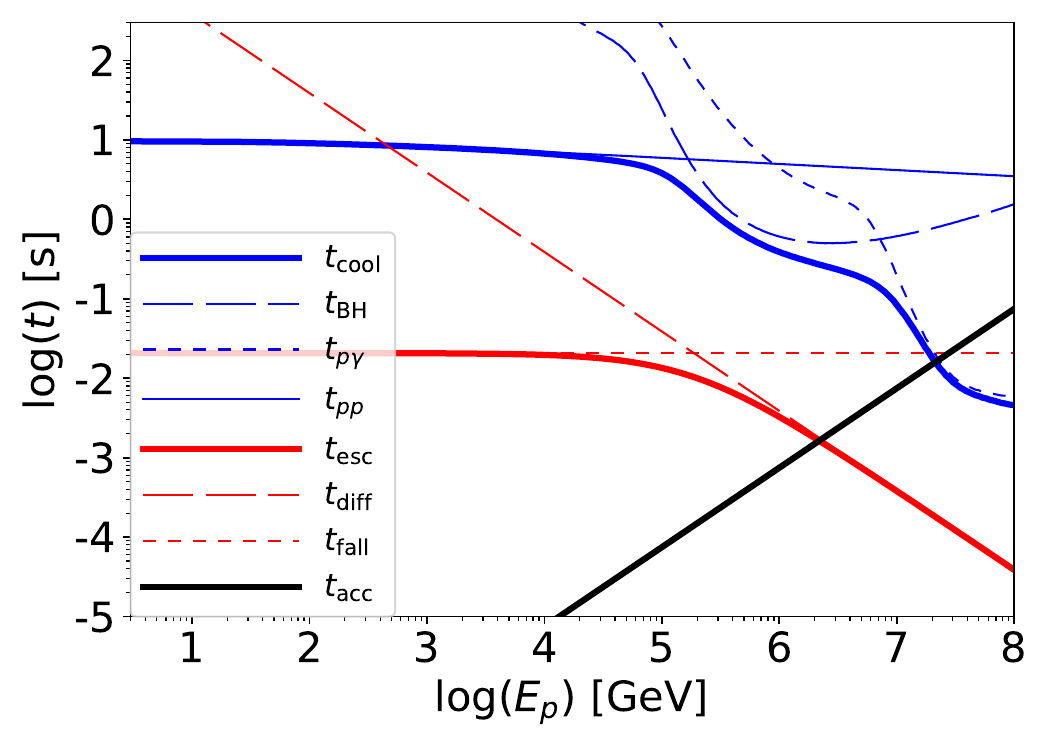}
    \caption{Various timescales as a function of proton energies for a typical IBH in a typical molecular cloud (top) and for parameters that can explain a LHAASO dark source, J0007+05659u (bottom). The solid-blue, solid-red, and solid-black lines represent the cooling, escape, and acceleration timescales, respectively. The thin-red-dashed and thin-red-dotted lines represent diffusion and inflall timescales, respectively. The thin-blue-solid, thin-blue-dashed, and thin-blue-dotted lines represent $pp$, $p\gamma$, and Bethe-Heitler cooling timescales, respectively. The parameters for both models are given in Table \ref{tab:param}.}
    \label{fig:times}
   \end{center}
  \end{figure}

\begin{table}[tbp]
  \caption{Parameter sets in our models. See Section \ref{sec:lhaaso} for values on $R_{\rm MC}$ and $B_{\rm MC}$. See Section \ref{sec:pevatron} for values on $M_\bullet$ and $V_k$. \label{tab:param}}
 Shared parameters\\
  \begin{tabular}{cccccccccc} \hline
    $\mathcal{R}$ & $\alpha$ & $\beta$ & $\lambda_w$ & $f_{\rm CR}$ & $\eta_{\rm rec}$ & $\eta_{\rm diff}$ & $s_{\rm{inj}}$\\ \hline
    10 & 0.3 & 0.1 & 0.1 & 0.035 & 10 & 10 & 2.0 \\ \hline
   \end{tabular}
\\
 Model parameters \\
    \begin{tabular}{c|cccccc} \hline
    Model & $M_\bullet$ & $n_{\rm MC}$ & $V_k$ & $R_{\rm MC}$ & $B_{\rm MC}$ &  $d$ \\
     & [$M_\odot$] & [cm$^{-3}$] & [km s$^{-1}$] & [pc] & [$\rm\mu G$] & [kpc]\\
     \hline
     Typical & 10 & 100 & 20 & 20 & 10 & 0.50 \\
     J0007 & 20 & 1000 & 20 & 5.0 & 30 & 2.0 \\
   \end{tabular}
  \label{tb_para}
\end{table}

We consider that MADs around IBHs accelerate CRs by magnetic reconnection. GRMHD simulations confirmed that MADs dissipates magnetic energy by magnetic reconnection \citep{BOP18a,2022ApJ...924L..32R}. A fraction of reconnection occurs in the relativistic regime where the magnetic energy density is higher than the rest-mass energy density of the plasma, i.e., $\sigma=B^2/(4\pi n_p m_pc^2)>1$. Such relativistic reconnection accelerates non-thermal particles very efficiently according to particle-in-Cell (PIC) simulations \citep{2001ApJ...562L..63Z,2020PhPl...27h0501G}. 

The non-thermal particles are subject to energy loss by cooling and escape processes. To obtain the spectrum of non-thermal protons in MADs, we solve the transport equation in steady state and one-zone approximations:
\begin{equation}
 \frac{d}{dE_p}\left(\frac{E_pN_{E_p}}{t_{\rm cool}}\right) -\frac{N_{E_p}}{t_{\rm esc}}+\dot{N}_{p,\rm inj} = 0,\label{eq:CRtransport}
\end{equation}
where $N_{E_p}$ is the number spectrum, $E_p$ is the proton energy, $t_{\rm cool}$ is the cooling time, $t_{\rm esc}$ is the escape time, and $\dot{N}_{p,\rm inj}$ is the injection term.

\citet{2023ApJ...956L..36Z} performed  3D PIC simulations, which indicated that the particle acceleration timescale by magnetic reconnection can be estimated as $t_{\rm acc}\approx \eta_{\rm rec}r_L/c$ , where $r_L=E_p/(eB)$ is the Larmor radius and $\eta_{\rm rec}\simeq10$ is the reconnection rate. Their simulations also show that the acceleration process forms a power-law spectrum of non-thermal particles with a spectral index of 2.
Based on this, we use the injection term of $\dot{N}_{p,\rm inj}= \dot{N}_0 (E_p/E_{p,\rm cut})^{-2}\exp(-E_p/E_{p,\rm cut})$, where $\dot{N}_0$ is the normalization factor and $E_{p,\rm cut}$ is the cutoff energy. We determine $\dot{N}_0$ by $L_{\rm CR} = \int\dot{N}_{p,\rm inj}E_pdE_p = f_{\rm CR}\dot{M}_\bullet c^2$. $E_{p,\rm cut}$ is determined by balancing the acceleration and loss timescales, i.e., $t_{\rm acc}=t_{\rm loss}$, where the loss timescale is given by $t_{\rm loss}^{-1}=t_{\rm esc}^{-1}+t_{\rm cool}^{-1}$.

As for escaping processes, we consider diffusive escape and infall to the IBH. We assume that the diffusive escape timescale is given by $t_{\rm diff}=3R_{\rm MAD}^2/(\eta_{\rm diff} r_L c)$, as in Bohm diffusion. The infall timescale is given by $t_{\rm fall}=R_{\rm MAD}/V_R$, where $V_R=(1/2)\alpha V_K$ is the radial velocity. The total escape rate is $t_{\rm esc}^{-1}=t_{\rm diff}^{-1}+t_{\rm fall}^{-1}$. Equating these two timescales, we obtain the escape energy above which the protons efficiently escape from the MAD: 
\begin{eqnarray}
 E_{p,\rm esc}= \frac{3eBR_{\rm MAD}V_R}{\eta_{\rm diff} c} \simeq 41 M_1n_{\rm MC,2}^{1/2}\lambda_{w,-1}^{1/2}V_{\rm eff,6.3}^{-3/2} \nonumber\\
\times\mathcal{R}_1^{-3/4}\alpha_{-0.5}^{1/2}\beta_{-1}^{-1/2}\eta_{\rm diff,1}^{-1}\rm~TeV.
\end{eqnarray}
For $E_p < E_{p,\rm esc}$, protons are mostly fall to the IBH. 

As for cooling processes, we consider proton synchrotron, $pp$ collision, photomeson production, and Bethe-Heitler processes. These are calculated using the same method as in \cite{2020ApJ...905..178K}. The target photons in MADs are computed with the same method as in \citet{2021ApJ...922L..15K}. Within the range of our interest, these processes are inefficient, compared to the escape processes as shown in Figure \ref{fig:times}.

Since diffusive escape limits the CR acceleration in this system, the cutoff energy, $E_{p,\rm cut}$, is determined by balancing the acceleration and diffusive escape timescales, which gives
\begin{eqnarray}
 E_{p,\rm cut}= eBR_{\rm MAD}\sqrt{\frac{3}{\eta_{\rm diff}\eta_{\rm rec}}}\simeq0.48 M_1n_{\rm MC,2}^{1/2}V_{k,6.3}^{-3/2}\nonumber\\
\times\lambda_{w,-1}^{1/2}\mathcal{R}_1^{-1/4}\alpha_{-0.5}^{-1/2}\beta_{-1}^{-1/2}\eta_{\rm diff,1}^{1/2}\eta_{\rm rec,1}^{1/2} \rm~PeV.\label{eq:Epcut}
\end{eqnarray}
Thus, MADs around IBHs can accelerate CR protons up to PeV energies if the system has low $V_k$, high $M_\bullet$, or high $n_{\rm MC}$.
This value is similar to the Hillas energy, $E_{\rm Hillas}=eBR_{\rm MAD}$, which is essentially determined by the energy budget, $\dot{M}c^2\sim10^{35}\rm~erg~s^{-1}$. Therefore, isolated black holes can achieve PeV energies even if we consider different scenarios, such as particle acceleration in jets \citep{2012MNRAS.427..589B,IMT17a}.

\section{IBHs in Molecular Clouds as LHAASO dark sources}\label{sec:lhaaso}

  \begin{figure}
   \begin{center}
    \includegraphics[width=\linewidth]{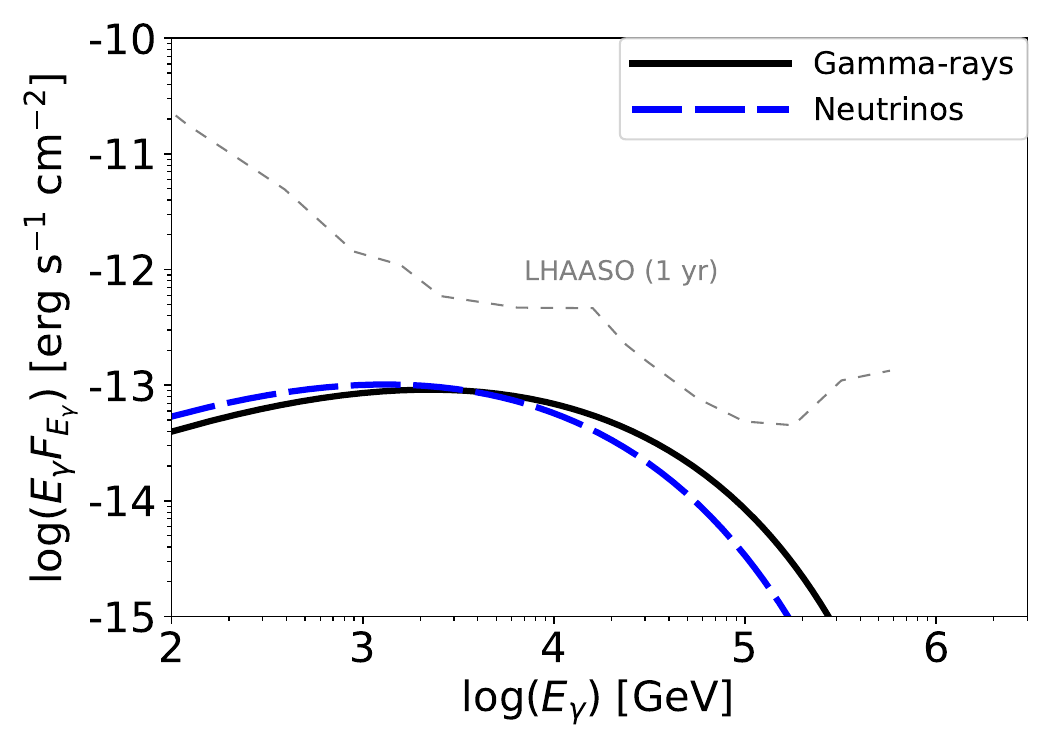}
    \includegraphics[width=\linewidth]{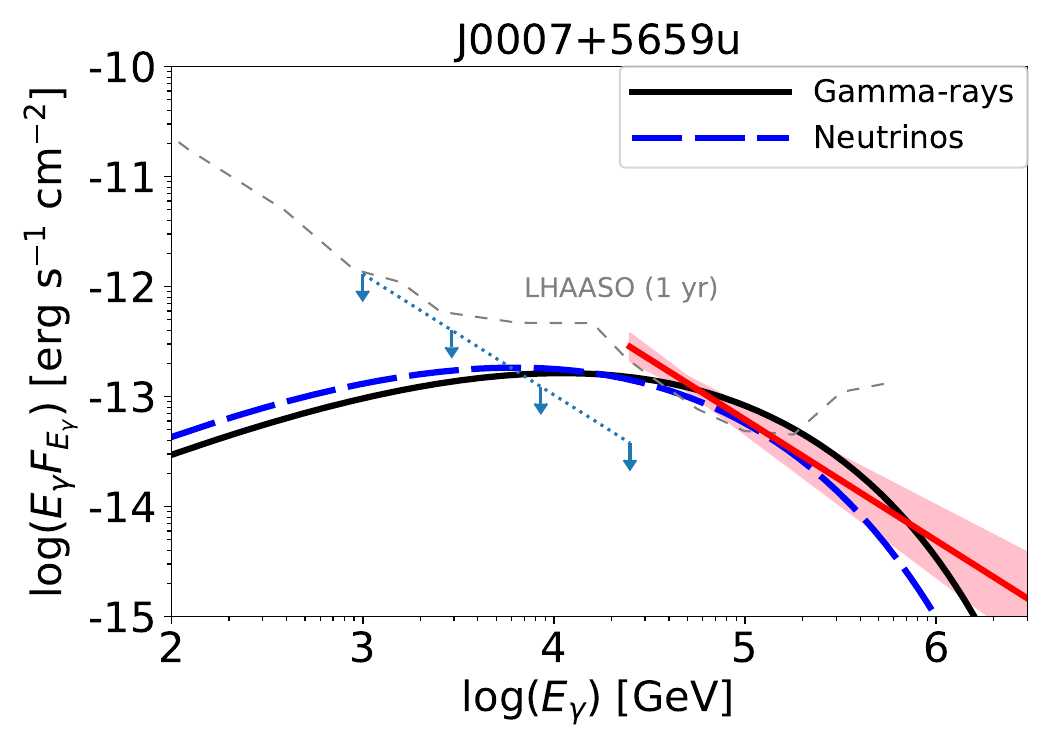}
    \caption{Gamma-ray spectra from molecular clouds that host IBHs. Top and bottom panels are for a typical case in a typical molecular cloud and for an optimistic case that matches a LHAASO dark source (J0007+5659u), respectively. Their parameter sets are tabulated in Table \ref{tab:param}. The thin-grey-dashed lines represent the LHAASO sensitivity \citep{2019arXiv190502773B}. The black-solid and blue dashed curves are our prediction on gamma-ray and neutrinos, respectively. The red line with a pink band and the thin-dotted line in the lower panel are the observed spectra and upper limit given in the first LHAASO catalog, respectively.}
    \label{fig:lhaaso_spe}
   \end{center}
  \end{figure}

  In this section, we discuss TeV-PeV gamma rays from a molecular cloud that hosts a wandering IBH. CR protons of $E>E_{p,\rm esc}$ escape from the MAD around the IBH. These CRs propagate in the host molecular cloud and interact with the ambient gas, leading to TeV-PeV gamma-ray emission via hadronuclear interactions before diffusively escaping from the molecular cloud.

 The injection rate of CRs to the molecular cloud is estimated by the escape term in Eq. (\ref{eq:CRtransport}), i.e., $\dot{N}_{\rm inj,MC}\approx N_{E_p}/t_{\rm diff}$, assuming that a molecular cloud hosts a single IBH.  We assume that the molecular cloud develops turbulence whose injection scale is the size of the molecular cloud, $R_{\rm MC}$. Assuming a Kolmogorov spectrum, the diffusion coefficient is approximated as \citep[e.g.,][]{2014PhRvD..89l3001H}
\begin{equation}
 D_{\rm MC}\approx \frac{\lambda_{\rm coh}c}{3}\left(\frac{4E_p^2}{E_{\rm coh}^2}+\frac{0.9E_p}{E_{\rm coh}}+\frac{0.23E_p^{1/3}}{E_{\rm coh}^{1/3}}\right),
\end{equation}
where $E_{\rm coh}=eB_{\rm MC}\lambda_{\rm coh}\simeq 0.18 B_{\rm MC,-5}R_{\rm MC,20}$ EeV, $\lambda_{\rm coh}=R_{\rm MC}/5$ is the coherence length, $e$ is the elementary charge, $B_{\rm MC}$ is the magnetic field strength in the molecular cloud, and $R_{\rm MC,20}=R_{\rm MC}/(20$ pc). The distribution of $B_{\rm MC}$ is given in \cite{2012ARA&A..50...29C}, where $B_{\rm MC}\sim1-10\rm~\mu G$ is obtained for a low density cloud (e.g., $n<3\times10^2\rm~cm^{-2}$). For a high density cloud, $B_{\rm MC}$ tends to be stronger. The diffusion timescale is given by
\begin{equation}
 t_{\rm diff,MC}=\frac{R_{\rm MC}^2}{2D_{\rm MC}}\simeq 7.1\times10^3 B_{\rm MC,-5}^{1/3}E_{\rm PeV}^{-1/3}R_{\rm MC,20}^{4/3}\rm~yr,
\end{equation}
where $E_{\rm PeV}=E_p/(1\rm~PeV)$ and we use $E_p\ll E_{\rm coh}$ for the last equation. 
Since the diffusive escape is the shortest loss process in this system\footnote{The crossing time of the molecular cloud is estimated to be $t_{\rm cross}=R_{\rm MC}/V_k\sim 9.8\times10^5 R_{\rm MC,20}V_{k,6.3}$ yr, which is much longer than $t_{\rm diff,MC}$.}, we can estimate the differential number density of CRs in the molecular clouds to be $N_{E_p}^{\rm MC}\approx \dot{N}_{\rm inj,MC}t_{\rm diff,MC}$. These CRs interact with gas in the molecular cloud via $pp$ inelastic collisions, whose cooling timescale is given by 
\begin{equation}
t_{pp,\rm MC}\approx \frac{1}{n_{\rm MC}\sigma_{pp}\kappa_{pp}c}\simeq 3.5\times10^5n_{\rm MC,2}^{-1}\rm~yr,
\end{equation}
where $\sigma_{pp}\simeq60$ mb and $\kappa_{pp}\sim0.5$ is the cross-section and inelasticity for pp collisions in relevant energies. These collisions produce neutral and charged pions which decay to gamma rays and neutrinos, respectively.
Since $\pi^\pm:\pi^0=2:1$ for $pp$ collisions, a third of the produced pions lead to gamma-ray production. The gamma-ray luminosity around 100 TeV can be roughly estimated as \citep[e.g.,][]{mal13,2017PTEP.2017lA105A}
\begin{eqnarray}
 L_\gamma&\approx& \frac13 f_{pp,\rm MC}f_{\rm bol}L_{\rm CR}
 \simeq 2.0\times10^{30}M_1^2n_{\rm MC,2}^2\lambda_{w,-1}\nonumber\\
&\times& V_{\rm eff,6.3}^{-3}B_{\rm MC,-5}^{1/3}R_{\rm MC,20}^{4/3}E_{\rm PeV}^{-1/3}f_{\rm bol,-1}\rm~erg~s^{-1},
\end{eqnarray}
where $f_{\rm bol}\approx1/(\ln(E_{p,\rm max}/\rm1~ GeV))\sim0.1$ is the bolometric correction factor and $f_{pp,\rm MC}=t_{\rm diff,MC}/t_{pp,\rm MC}$ is the pion production efficiency in the molecular cloud. This value is so low that it is challenging to detect gamma-rays even if the IBH is situated in a nearby molecular cloud at a distance of $\sim500$ pc from Earth, whose gamma-ray flux would be $\sim 1\times10^{-13}\rm~erg~s^{-1}~cm^{-2}$. Nevertheless, we can consider a wide range of $n_{\rm MC}$, $V_k$, and $M_{\rm BH}$, and some parameter sets could enhance the gamma-ray luminosity by a few orders of magnitude.

We numerically calculate the gamma-ray spectrum from the molecular clouds using the method by \cite{kab06} with the updated $pp$ cross-section given in \citet{2014PhRvD..90l3014K}. The results are shown in Figure \ref{fig:lhaaso_spe}. As seen in the top panel, we cannot expect gamma-ray detection if we use the typical molecular cloud parameters given in Table \ref{tab:param}. On the other hand, if we take an optimistic parameter set (see J0007 on Table \ref{tab:param}), the resulting gamma-ray emission can be luminous enough to be detected by LHAASO, as shown in the bottom panel. In our scenario, only high-energy CRs can escape from the MADs around IBHs. Thus, we do not expect GeV-TeV gamma-rays from the molecular cloud. This feature is consistent with the gamma-ray data of a LHAASO dark source, J0007+5659u. Here, we choose J0007+5659u because it is a peculiar object;  It is detected only by KM2A, detected at $E_\gamma > 100$ TeV, and located in the Galactic plane. Only two dark sources satisfy these conditions. Our model can explain the other dark source, J1959+1129u, with a similar parameter set, because their spectral and morphological features are similar to each other.  

In our scenario, we need to consider optimistic parameter sets for IBHs to make sources detectable by LHAASO. This is because with a typical parameter set, the cutoff energy of gamma rays are smaller than a few tens of TeV. The sensitivity curve of LHAASO exhibits a minimum at approximately 100 TeV, and typical IBHs in typical molecular clouds cannot emit such high energy photons. On the other hand, IBHs with the optimistic parameter set can emit $\sim100$ TeV gamma rays, enabling LHAASO to detect such systems even if they are located at several times more distant than the nearest molecular clouds. Because of their rarity, the nearest IBH detectable by LHAASO could be located at a few kpc away from the Earth. In this situation, the angular size of the molecular cloud is $\sim0.1$ deg, which is consistent with the size of the dark sources ($<0.18$ deg for J0007+5659u) reported by the LHAASO Collaboration \citep{2024ApJS..271...25C}.

Some of the dark sources, J0206+4302u and J0212+4254u, are located at high galactic latitude ($b=-17$ deg; \citealt{2024ApJS..271...25C}). Although typical giant molecular clouds are concentrated on the Galactic plane, dense gas clouds exist even in such a high galactic latitude \citep[e.g.,][]{NS16a,2019A&A...624A...6Y}. Quantitative evaluation whether our model can explain these sources are left as future work.

\section{Contribution of IBHs to PeV CRs on Earth }\label{sec:pevatron}

  \begin{figure}
   \begin{center}
    \includegraphics[width=\linewidth]{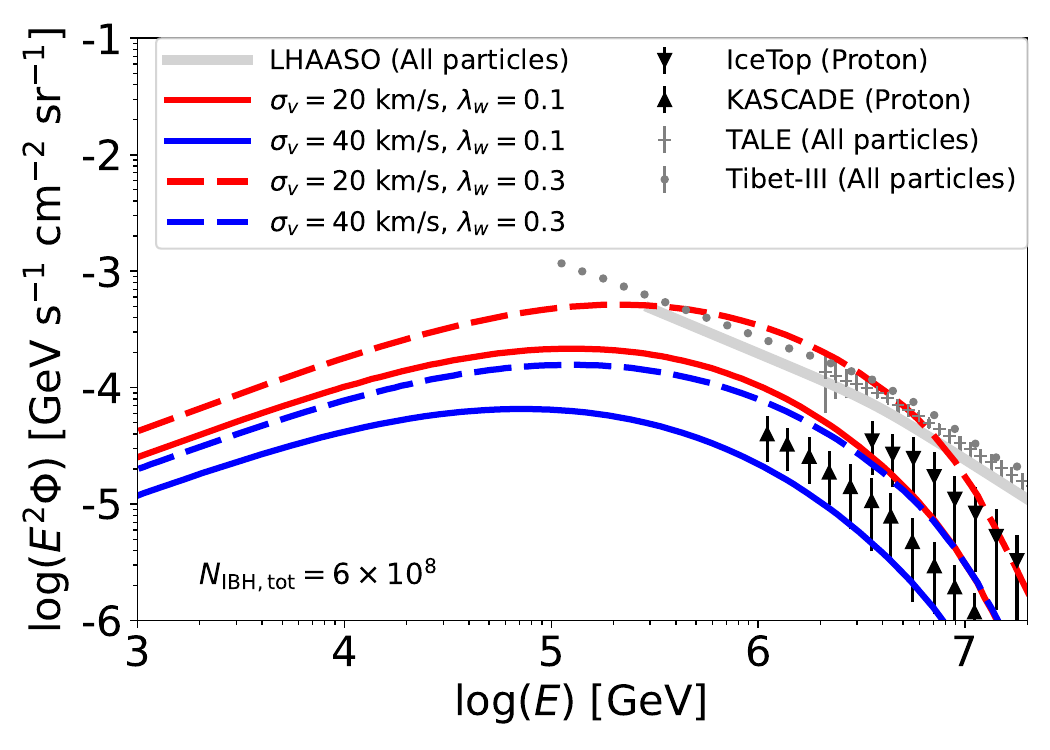}
    \caption{Comparison of our model prediction to the observed CR intensity on Earth. The red line represents our prediction. The data points are from KASCADE \citep{KASCADE13a}, IcethTop \citep{2019PhRvD.100h2002A}, TALE \citep{TA18a}, Tibet-III \citep{Amenomori:2008aa}, and LHAASO \citep{LHAASO:2024knt}. }
    \label{fig:crspe}
   \end{center}
  \end{figure}

In this section, we estimate contribution of IBHs in molecular clouds to PeV CRs observed on Earth. 
Both IBHs and molecular clouds should be concentrated on the inner part of our Galaxy. 
The distribution of the molecular gas in our Galaxy are given in \cite{NS16a}, which is concentrated within $\lesssim1-2$ kpc from the Galactic center. We estimate the volume filling factor of molecular gas in the Galactic center following the method of \cite{2018MNRAS.477..791T}, where the volume filling factor of molecular clouds, $\xi_{\rm MC}$, depends on galactocentric radius, $R_{\rm gc}$. We find that the volume filling factor in the inner Galaxy is $\xi_{\rm MC}\simeq0.02$ for $R_{\rm gc}\lesssim 1-2$ kpc, which is more than an order of magnitude higher than that of the solar neighborhood \citep{2000eaa..bookE2636B}. 
There should be density distribution within the molecular gas phase, and the higher density regions should have a smaller volume filling factor. We assume $d\xi/dn_{\rm MC}\propto n_{\rm MC}^{-2.8}$ following the previous work \citep{IMT17a,2018MNRAS.477..791T}.
  
Next, we describe the IBH distribution in our Galaxy. If the IBHs are formed by the evolution of the disk stars, the surface density distribution of IBHs should roughly follow the stellar distribution in the Galactic disk. The surface density profile of the disk component is given by the exponential function, $\Sigma_{\rm IBH}\sim \Sigma_0\exp(-R_{\rm gc}/R_d)$, where $R_d=2.15$ kpc and $\Sigma_0$ is the normalization factor \citep{2015ApJ...806...96L}. The total number of IBHs in our Galaxy is normalized by $N_{\rm IBH,tot}=2\pi \int dR_{\rm gc}\Sigma_{\rm IBH} R_{\rm gc}$. We set $N_{\rm IBH,tot}=6\times10^8$ \citep[e.g.,][]{2020ApJ...905..121A}, although this value has a large uncertainty. The total number of IBHs embedded in molecular clouds is estimated to be $N_{\rm IBH,MC}\approx\int dR_{\rm gc} 2\pi R_{\rm gc}\Sigma_{\rm IBH}\xi_{\rm MC}(H_{\rm MC}/H_{\rm IBH})$, where $H_{\rm MC}\sim0.075$ kpc and $H_{\rm IBH}$ are the scale heights of the molecular gas and IBHs, respectively. We assume $H_{\rm IBH}=0.3$ kpc, based on numerical computation for IBH distribution in our Galaxy \citep{2018MNRAS.477..791T}.

The velocity distribution of the IBH population, $\sigma_v$, is affected by the natal kick distribution. The Galactic distribution for BH X-ray binaries suggests that a fraction of BHs experienced a strong natal kick of $\gtrsim100\rm~km~s^{-1}$, but the majority of BHs are consistent with a weak natal kick of $V_k\sim10-50\rm~km~s^{-1}$ \citep{RIN17a,2024arXiv241116847N}.
Also, the discovery of an IBH by microlensing event also favors a lower value of $V_k<100\rm~km~s^{-1}$ \citep{2022ApJ...933...83S,2024ApJ...973....5K}.
Here, we assume that the kick velocity of the formation of IBHs are weak and the velocity dispersion of the IBH population is similar to that of the disk stars, i.e., $\sigma_v\sim20\rm~km~s^{-1}$.
We assume that the velocity distribution is given by a Gaussian with $\sigma_v$,
and the mean velocity of IBHs is $\sqrt{\pi/2}\sigma_v\sim25\rm~km~s^{-1}$. 
As for the mass distribution of IBHs, we use the mass distribution obtained by gravitational wave observations, which can be approximated as $dN/dM_\bullet\propto M_\bullet^{-3.5}$ within the range of $10~M_\odot\lesssim M_\bullet\lesssim 50 M_\odot$ \citep{KAGRA:2021duu}\footnote{Although the mass distribution of merging BHs are not represented by a power-law form, we use a single power-law mass distribution for simplicity. In addition, the minimum and maximum masses of the stellar-mass BH population are not well constrained by the GW data. }.
  
We use the leaky-box approximation to estimate the CR intensity on Earth. Using the distributions of parameters ($dN/dM_\bullet$, $dN/dV_k$, $d\xi/dn_{\rm MC}$), the CR injection rate from IBHs to ISM is estimated as
\begin{eqnarray}
 E_p Q_{E_p}&\approx& \int dM_\bullet \int dn_{\rm MC} \int dV_k N_{\rm IBH,MC} \\
&\times& E_p^2\dot{N}_{\rm inj,MC}\frac{d\xi}{dn_{\rm MC}}\frac{dN}{dM_\bullet}\frac{dN}{dV_k},\nonumber
\end{eqnarray}
where we normalize the distribution by $\int dM_\bullet dN/dM_\bullet=1$, $\int dV_k dN/dV_k=1$, and $\int dn_{\rm MC}d\xi/dn_{\rm MC}=1$. Here, we assume that $dN/dM_\bullet$, $dN/dV_k$, and $d\xi/dn_{\rm MC}$ are independent of $R_{\rm gc}$ for simplicity.
The confinement timescale of the CR protons in our Galaxy is estimated by using the grammage, $X_{\rm esc}$, which indicates the amount of matter in the CR path length from the source to the Earth. Based on recent experiments, the grammage is estimated to be $X_{\rm esc}\simeq2.0(E_p/250\rm~GeV)^{-\delta}$, where $\delta=0.46$ for $E_p<250$ GeV and $\delta=0.33$ for $E_p>250$ GeV \citep{PAMELA14a,AMS16a}. Balancing the injection from IBHs and escape from the ISM, the CR proton intensity on Earth is estimated as \citep[e.g.,][]{KMM18a,2019PhRvD..99f3012M}
\begin{equation}
 E_p^2\Phi_p\approx \frac{E_pQ_{E_p}X_{\rm esc}}{4\pi M_{\rm gas}},
\end{equation}
where $M_{\rm gas}\simeq8\times10^9M_\odot$ is the total gas mass in our Galaxy \citep{NS16a}. 

The resulting CR proton spectrum is shown in Figure \ref{fig:crspe}. We find that IBHs in molecular clouds could provide a significant contribution to the PeV CRs observed on Earth. Typical IBHs in typical density molecular clouds can accelerate CRs up to $\lesssim1$ PeV. On the other hand, IBHs with high $M_\bullet$, low $V_k$, and high $n_{\rm MC}$ can accelerate CRs up to 1-10 PeV (see Equation (\ref{eq:Epcut})), which enables IBHs in molecular clouds to contribute to the super-knee CR component.

Although our scenario can explain the PeV CR data with a reasonable parameter set, it contains uncertain parameters, such as the total number of IBHs, $N_{\rm IBH,tot}$, the reduction factor of the accretion rate, $\lambda_w$, and the velocity dispersion of the IBH, $\sigma_v$. We calculate the CR intensities with various set of parameters, which are shown in Figure \ref{fig:crspe}. The CR intensities at PeV energies are higher for higher $\lambda_w$ and lower $\sigma_v$. The intensity is also proportional to $N_{\rm IBH,tot}$. These parameters could be constrained by future observations or simulations. Especially,  $N_{\rm IBH,tot}$ and $\sigma_v$ will be obtained by wide and deep optical surveys, such as LSST and Roman, because these surveys would be able to identify multiple IBHs by microlensing events \citep[e.g.,][]{2018arXiv181203137S}.  

The discrepancy between KASCADE and IceTop likely originates from the uncertainty of the hadronic interaction models. The recent LHAASO result indicates that the mass composition around the knee energy is dominated by light elements \citep{LHAASO:2024knt}, which is consistent with the IceTop result. Our reference model predicts that the proton contribution is 30\% of the observed knee energies of 4 PeV. If we use slightly higher $\lambda_w$ or lower $\sigma_v$, our model prediction would be consistent with the LHAASO and IceTop results.   

\section{Summary \& Discussion}

We propose that IBHs in molecular clouds can be the origin of LHAASO dark sources and PeV CRs observed on Earth. IBHs accrete gas in molecular clouds, which lead to the formation of MADs around IBHs. In the MADs, CR protons can be accelerated up to PeV energies via magnetic reconnection in the vicinity of IBHs. Then, these PeV CRs escape from the MADs and propagate in the ambient molecular clouds, which leads to gamma-ray emission from the clouds via hadronuclear interactions. This gamma-ray signals can explain LHAASO dark sources, from which we observe 100 TeV photons without GeV-TeV gamma-ray counterparts. The vast majority of the PeV CRs escape from the molecular clouds and are injected into the ISM in our Galaxy. These PeV CRs can provide a significant contribution to the PeV CR intensity observed on Earth with a reasonable parameter set.

Based on our scenario, the dark sources detected by LHAASO should be associated with dense clouds. Obvious associations are currently not reported (but see \citealt{Xie:2024bbu} for a tentative association with a small, nearby molecular cloud around J0007+5659u), despite that radio Galactic plane surveys have been already conducted \citep[e.g.,][]{2001ApJ...547..792D,2017PASJ...69...78U}. The LHAASO angular resolution is larger than the typical field-of-view of radio telescopes, and our scenario demands relatively distant and denser gas associated with the dark LHAASO sources, both of which make the identification of dense gas clouds challenging. Improvements for angular resolution of LHAASO and hihg-sensitivity radio surveys with high-density tracers are necessary to identify a dense cloud or rule out the existence of it.

\citet{2016Natur.531..476H} reported the detection of 100-TeV gamma-rays from the central molecular zone, suggesting the existence of PeVatron at Galactic Center. Our scenario would naturally explain the existence of PeVatron at Galactic Center because both IBHs and molecular clouds are concentrated on the Galactic Center region as discussed in Section \ref{sec:pevatron}.  

Based on our scenario, we have $\sim20$ IBHs embedded in molecular clouds within 1 kpc from the Earth, regardless of the LHAASO detectability. Although we have molecular clouds as close as $\sim500$ pc, typical IBHs embedded in typical molecular clouds cannot emit gamma-rays detectable by LHAASO (see Figure \ref{fig:lhaaso_spe}). Luminous gamma-ray signals demand high $M_\bullet$, low $V_k$, high $n_{\rm MC}$, or high $B_{\rm MC}$, which are likely to be achieved in the inner Galaxy except for low $V_k$. Since the systems satisfying these conditions are rare, our scenario can be consistent with the current LHAASO data. The number of IBHs detectable by LHAASO depends on the mass and magnetic field distributions in molecular clouds \citep[e.g.,][]{2012ARA&A..50...29C,2017ApJ...836..175K,2018PASJ...70S..59K}, and the detailed estimate on this is left as a future work.

The diffusion coefficient of CRs in molecular clouds also has some uncertainty. Since the IBH will provide a lot of CRs into molecular clouds, CR streaming will lead to current driven instabilities \citep[e.g.,][]{Ski75a,Bel04a}, causing an efficient confinement of CRs in molecular clouds. In this case, the gamma-ray signals would be stronger than that given in our scenario. On the other hand, the low-ionization rate in molecular clouds might suppress the streaming instability \citep{2007A&A...475..435R,2021MNRAS.504.2405A}, which could lead to more efficient diffusion. In this case, the gamma-ray signals would be similar to that in our scenario. 

Multi-wavelength observations are useful to test our scenario. The optical and soft X-ray emissions from the IBH should be strongly attenuated due to the dust and gas in the molecular cloud. The column density of a typical cloud is estimated as $N_H\approx R_{\rm MC}n_{\rm MC}\sim 6\times10^{21}R_{\rm MC,20}n_{\rm MC,2}\rm~cm^{-2}$ \citep[e.g.,][]{2022A&A...666A.165S}. Thus, soft X-ray ($\lesssim1$ keV) and optical photons $(\gtrsim4\times10^{14} \rm~Hz)$ should be strongly attenuated \citep[e.g.,][]{2000ApJ...542..914W,1989ApJ...345..245C,2009MNRAS.400.2050G}. On the other hand, hard X-rays ($\gtrsim1$ keV) and infrared photons ($\lesssim4\times10^{14}$ Hz) do not suffer from attenuation, and thus, follow-up observations to LHAASO dark sources by hard X-ray and mid-infrared telescopes may be able to identify IBHs in molecular clouds. In addition, IBHs in molecular clouds are expected to emit GeV gamma-rays via curvature radiation and inverse Compton scattering from BH magnetospheres \citep{2016ApJ...833..142H,2024ApJ...964...78K,2025arXiv250209181K}.
Since molecular clouds are concentrated on the central part of our Galaxy, UHE gamma-ray detectors in the southern hemisphere, such as ALPACA \citep{2024arXiv241214550A} and SWGO \citep{Abreu:2019ahw}, will increase the number of dark sources. Details of the multi-wavelength observation strategy are planned and to be investigated in the future.
The protons accelerated in the MAD can also emit gamma-rays via proton synchrotron or $pp$ inelastic collisions in the MAD \citep{2025arXiv250117467K}. However, they are typically fainter than the GeV emission from the magnetosphere and TeV-PeV emission by $pp$ inelastic collisions in the molecular cloud. 

Future neutrino observations may also provide a robust test on our scenario, because hadronic gamma-ray sources are accompanied with neutrinos \citep[e.g.,][]{mal13,2023PhRvD.107d3002S}. Neutrino emission from these sources are challenging to be detected by current neutrino neutrino detectors. As shown in Figure \ref{fig:lhaaso_spe}, our model predicts that the neutrino flux is expected to be $\sim10^{-13}\rm~erg~s^{-1}~cm^{-2}$ at 10--100 TeV, but this neutrino flux is an order of magnitude lower than the sensitivity of IceCube \citep{Aartsen:2019fau}. If we identify more dark sources, stacking analyses could potentially confirm or constrain our scenario. Also, these neutrino signals will be detectable by future neutrino detectors, such as KM3NeT \citep{KM3NeT16a}, IceCube-Gen2 \citep{Aartsen:2020fgd}, TRIDENT \citep{2023NatAs...7.1497Y}, P-ONE \citep{P-ONE:2020ljt}, and HUNT \citep{Huang:2023mzt}. Future neutrino observations, together with UHE gamma-ray detectors, will be able to unravel the nature of cosmic PeVatrons near future.

\begin{acknowledgments}
S.S.K. thank Sei Kato and Dmitry Khangulyan for useful discussion and comments. This work is supported by KAKENHI No. 22K14028, 21H04487, 23H04899 (S.S.K.), 22K14080 (M.I.N.K), 22K0043 (K.T.), the Graduate Program on Physics for the Universe (GP-PU), Tohoku University (K.K.), JST SPRING, Grant Number JPMJSP2114 (K.K.). S.S.K. acknowledges the support by the Tohoku Initiative for Fostering Global Researchers for Interdisciplinary Sciences (TI-FRIS) of MEXT's Strategic Professional Development Program for Young Researchers.
\end{acknowledgments}

%

\vspace{5mm}








\bibliography{sample631}{}
\bibliographystyle{aasjournal}



\end{document}